\documentclass[prl,aps,showpacs,preprint,superscriptaddress]{revtex4-1}

\usepackage{graphicx}
\usepackage{bm}
\usepackage{amsfonts}
\usepackage{amsmath}
\usepackage{amssymb}
\usepackage{color}

\newcommand{\bu}{\boldsymbol{u}}

\newcommand{\De}{\boldsymbol{\Delta}}
\newcommand{\xp}{\boldsymbol{x}_\mathrm{i}}
\newcommand{\vp}{\boldsymbol{v}_\mathrm{i}}
\newcommand{\de}{\mathrm{d}}
\newcommand{\dd}[2]{\frac{\de{#1}}{\de{#2}}}

\begin{document}

\title{ How Violent are the Collisions of Different Sized Droplets in a Turbulent Flow?}

\author{Martin James} \affiliation{Department of Physics, Indian Institute of Science, 
Bangalore 560012, India}\email{martin.nilambur@gmail.com}
\author{Samriddhi Sankar Ray} \affiliation{International Centre for
  Theoretical Sciences, Tata Institute of Fundamental Research,
  Bangalore 560089, India}\email{samriddhisankarray@gmail.com}

\begin{abstract}
We study the typical collisional velocities in
a polydisperse suspension of droplets in two and three-dimensional turbulent
flow and obtain precise theoretical estimates of
the dependence of the impact velocity of particles-pairs on their relative
sizes. These analytical results are validated against data from our direct numerical
simulations. We show that the impact velocity saturates exponentially with the
inverse of the particle-size ratios. Our results are important to
model  coalescence or fragmentation (depending on the impact velocities) and
will be crucial, for example, in obtaining precise coalescence kernels to
describe the growth of water droplets which trigger rain in warm clouds.
\end{abstract}


\maketitle

A wide variety of industrial, experimental and natural phenomena -- such as
volcanic eruptions, embryonic planetesimals in circumstellar disks and growth
of water droplets triggering rain in warm clouds -- involves coalescence,
collisions, and fragmentation of small, heavy particles carried by a turbulent flow~\cite{intro-ref}.
Given the ubiquitousness and generality of particle-laden flows, problems of
this class have been the subject of intense research in the last few years. 
As a result significant progress has been made in our
understanding of not only the appropriate theoretical~\cite{MR}, numerical~\cite{bec-numerics} 
and mathematical framework~\cite{bec-math} to study such systems but also the nature of coalescence
and collisions in both idealised~\cite{idealised} and realistic settings~\cite{coal}. 
In the last couple of years, for example, several authors have elucidated, theoretically and in
experiments, questions related to the distribution of relative velocities of
approaching particles ~\cite{mehlig,saw} and the extreme events
associated with them. Such studies are a key building block in
helping us to eventually model effective collision kernels which would be
crucial in computing, for example, droplet distributions in warm clouds.

Much of the work related to the issue of approach rates and relative velocities
of particle-pairs deal with systems of identical particles~\cite{bec2005,bec2010,bec2014}.
However in nature, the distribution of particles is typically inhomogeneous.
Hence particles of different sizes interact with one another. Furthermore, to
critically understand if coalescences dominate -- leading to the growth of
larger and larger droplets -- an estimate of the strength of impact velocities
of particles with different radii is crucial. In particular the role of
preferential concentration (which changes with particle radii) of particles at
small scales as well as possible large velocity differences due to
caustics~\cite{caustics} and the sling effect~\cite{sling} ought to
result in non-trivial collisional velocities in an inhomogeneous size
distribution of particles in a turbulent flow.

Given all of this, it behooves us to ask the rate at which droplets or
aggregates grow in an inhomogeneous size distribution of particles which are carried
by the same turbulent flow. This, it is worth stressing, is the natural setting
for processes which are of relevance in nature and in the most general of
laboratory settings. A first step in this direction is to understand how
violent the impact velocities amongst different particles are in such
polydisperse suspensions. A quantitative measure of this, in turn, will
suggest whether the impact between particle-pairs of radii $a_1$ and $a_2$ will
result in a coalescence -- leading to a larger droplet -- or fragmentation~\cite{frag}.
Thus,  in this paper we  address (in two and three dimensions), 
numerically and theoretically, the critical question of the dependence of the 
impact velocities of particle-pairs on the relative sizes of the particles.  

\begin{table*}
\begin{center}
\begin{tabular}{@{\extracolsep{\fill}} |c|c|c|c|c|c|c|c|c|c|c|}
\hline
Dimension & $N$ & $N_p$ & $\nu$ & $k_{\rm inj}$ & $\eta$ & $k_{\rm max}\eta$ & $\lambda$ & $Re_{\lambda}$ & $\tau_\eta $ \\ 
\hline \hline
3D & 512 & $10^6$ & 0.001 & 1 and 2 & 0.0059 & 1.01 & 0.0856 & 121 & 0.0351 \\ 
\hline
2D & 1024 & $10^5$ & $10^{-5}$ & 4 & 0.0044 & 1.50 & 0.20 & 440 & 1.92 \\ 
\hline
\end{tabular}
\end{center}
\caption{ Parameters for our simulations: $N$ is the number of grid points
along each direction, $N_p$ is the number of Lagrangian and heavy inertial particles, 
$\nu$ the kinematic viscosity, $\epsilon$ is the fixed energy input, $k_{\rm inj}$ 
the forcing wavenumber, $\eta \equiv (\nu^3/\varepsilon)^{1/4}$ the dissipation scale, 
$\lambda\equiv \sqrt{\nu E/\varepsilon}$ the Taylor microscale, 
$Re_{\lambda}\equiv u_{\rm rms} \lambda/\nu$ the Taylor-microscale Reynolds number,
and $\tau_\eta \equiv \sqrt{\nu/\varepsilon}$ the Kolmogorov time scale.}
\label{table1}
\end{table*}

We consider a fluid flow whose velocity $\bm u$ is a solution to the
incompressible Navier--Stokes equation
\begin{equation}
  \partial_t \bu + (\bu\cdot\nabla)\bu = -\nabla p + \nu\nabla^2
  \bu +\boldsymbol{f}, \quad \nabla\cdot\bu = 0,
  \label{eq:navier-stokes_3d}
\end{equation}
where $\nu$ designates the fluid kinematic viscosity. In two dimensions (2D), it is often 
convenient to re-write this in the vorticity ($\omega$)-stream function ($\psi$) 
formulation~\cite{prasad} as 
\begin{equation}
\partial_t \omega - J(\psi,\omega) = \nu \nabla^2 \omega + f_{\omega}
- \mu \omega ,
\label{eq:navier-stokes_2d}
\end{equation}
where $J(\psi,\omega) \equiv (\partial_x \psi)(\partial_y \omega) - (\partial_x
\omega) (\partial_y \psi)$ and $\mu$ is the
coefficient of Ekman friction. At the point ($x$,$y$) the velocity $\bm u
\equiv (-\partial_y \psi, \partial_x \psi) $ and the vorticity $\omega = \nabla
\psi$. 

We now study the dynamics of small inertial particles (droplets) which are 
suspended in a turbulent flow field obtained as a solution of Eq.~\eqref{eq:navier-stokes_3d} in 
three dimensions (3D) or of Eq.~\eqref{eq:navier-stokes_2d} in 2D in the limit of small 
$\nu$ or large Reynolds numbers. We assume that our particles are much smaller 
than the Kolmogorov scale $\eta$, much heavier than
the surrounding fluid, and with a small Reynolds number associated to their
slip velocity. The motion of the i-th particle, in the turbulent fluid, is damped 
through a viscous Stokes drag and their trajectories $\xp(t)$ are defined via 
\begin{equation}
  \dd{\xp}{t} = \vp,\quad \dd{\vp}{t} =
  -\frac{1}{\tau_\mathrm{p}}\left[ \vp-\bu(\xp,t)\right].
  \label{eq:particles}
\end{equation}
The relaxation time  $\tau_\mathrm{p} \!=\!
2\rho_\mathrm{p}a^2/(9\rho_\mathrm{f}\nu)$, where $\rho_\mathrm{p}$ and
$\rho_\mathrm{f}$ are the particle and fluid mass density respectively and $a$ the
particle radius, allows us to define a non-dimensional Stokes number 
$St = \tau_p/\tau_\eta$; the small time-scale $\tau_\eta$ is the 
Kolmogorov time scale and an intrinsic property of the fluid~\cite{frisch95}. 

We perform direct numerical simulations (DNSs) of Eq.~\eqref{eq:particles}
coupled with Eq.~\eqref{eq:navier-stokes_3d} in 3D or Eq.~\eqref{eq:navier-stokes_2d}
in 2D. The Navier-Stokes equations are solved in a 2$\pi$ periodic domain (with
the number of collocation points $N^3$ (3D) or $N^2$ (2D)) by using a standard
pseudospectral method and a second-order Runge-Kutta scheme for time-marching.
We maintain statistically steady homogeneous, isotropic turbulence via the
large-scale forcing $\boldsymbol{f}$ in 2D and by a constant energy injection
in 3D on wavenumbers $k_{\rm inj}$. The parameters of both sets of simulations are given 
in Table~\ref{table1}.

\begin{figure}[h]
\includegraphics[width=\columnwidth]{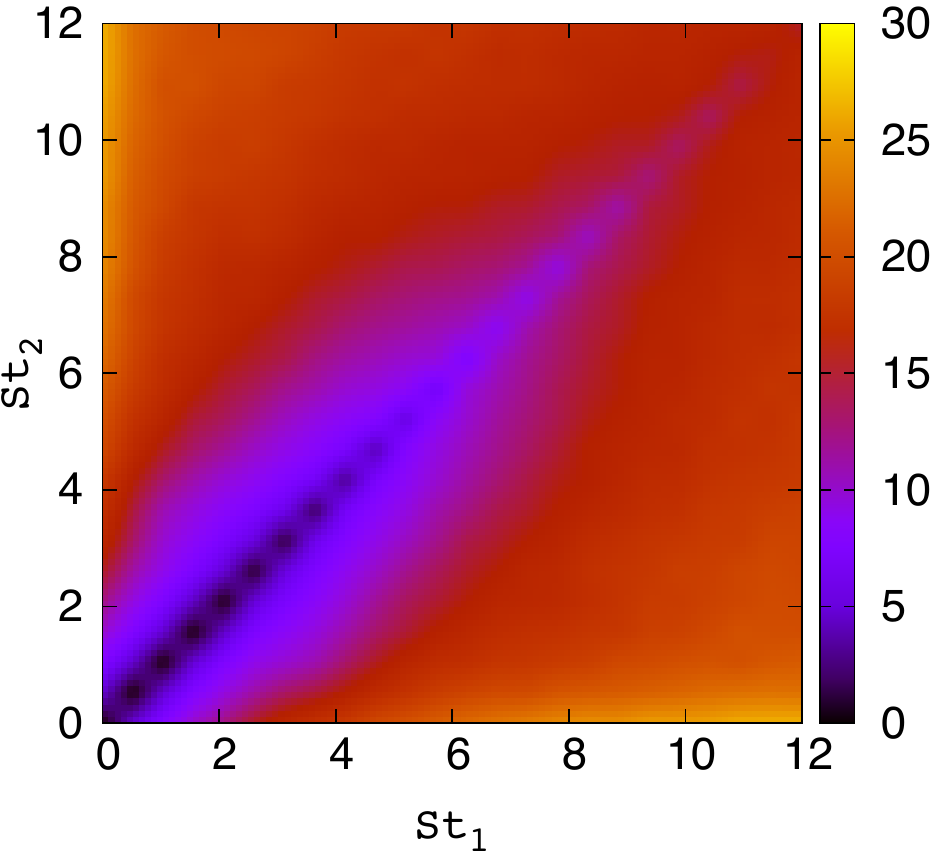}
\caption{(color online) Impact velocity $\De(St_\mathrm{1}, St_\mathrm{2})$ as a 
function of Stokes numbers of approaching particles in 2D.}
\label{fig:heatmap}
\end{figure}

We solve Eq.~\eqref{eq:particles} numerically for $N_p$ (see Table~\ref{table1})
non-interacting particles associated with a Stokes time $\tau_p$; we choose 
50 different $\tau_p$ in 2D and 20 different $\tau_p$ in 3D.
We thus integrate individual particle trajectories for different
values of $\tau_\mathrm{p}$ and the fluid velocity $\bm u$ at their location is
evaluated by linear interpolation. We study the dynamics of the particles within the 
framework of {\it ghost collisions} and hence we do not perform any real collisions; results 
of the effect of collisions will be reported elsewhere~\cite{martin}.
Given the assumption on the particle sizes, we work within the framework of one-way coupling 
between the particles and the fluid and ignore any effect of the particles on the flow.

\begin{table}
\begin{center}
\begin{tabular}{@{\extracolsep{\fill}} |c|c|c|c|c|}
\hline
Case & $St_1$ & $St_2$ & Prediction & Figure\\
\hline \hline
Case 1 & -- & $St_1$ & $\De = \De_0\exp{(-\tau/St)}$ & Fig.~\ref{fig:same}\\
\hline
Case 2 & $St_1 \ll 1$ & $St_2 \lesssim 1$ & $\De \sim St_2$ & Fig.~\ref{fig:s_l}\\
\hline
Case 3 & $St_1 \ll 1$ & $St_2 \gg 1$ & $\De = \De_0\exp{(-\tau/ St_2)}$ & Fig.~\ref{fig:s_l} \\
\hline
Case 4 & $St_1 \gtrsim 1$ & $St_2 \neq St_1$ & None & Fig.~\ref{fig:l_s} \\
\hline
\end{tabular}
\end{center}
\caption{A summary of the different asymptotics and the theoretical predictions for $\De$.}
\label{cases}
\end{table}

The impact velocity between colliding droplets, which determines the chance of
coalescence or fragmentation, is defined as the velocity difference of the two
particles at vanishing separation. Thence, we define the impact velocity
between two particles, labelled 1 and 2, of Stokes numbers $St_1$ and $St_2$ as
$\De =  \langle | ({\bf v}_2 - {\bf
v}_1)\cdot\hat {\bf r} | \rangle$, where the unit vector $\hat {\bf r}$ defines
the vector connecting the centers of the two particles, with the constraint
that $({\bf v}_2 - {\bf v}_1)\cdot\hat {\bf r} < 0$ which define a pair of
approaching particles~\cite{foot1}. The averaging $\langle \cdot \rangle$ is 
defined over all colliding pairs. 

In Fig.~\ref{fig:heatmap} we show a pseudo-color plot of the amplitude of the
impact velocity $\De$  as a function of the Stokes numbers of the colliding
particles is shown from our data from the 2D simulations, with similar results
obtained in our 3D DNS. (In this plot and all subsequent plots, we show $\De$
normalised by the Kolmogorov velocity $u_\eta = \eta/\tau_\eta$ of the
underlying turbulent fluid.) Qualitatively, it is easy to understand the
diagonal ($St_1 = St_2$) behaviour of $\De$: The two-particle velocity
correlation between particles (when the size of one particle is fixed and the
other varied) attains a maximum when particles are of the same size ($St_1 =
St_2$)~\cite{Wang}.  Consequently,  $\De$ attains a minimum when the
approaching particles have the same Stokes number as is clearly seen in
Fig.~\ref{fig:heatmap}. For larger Stokes numbers, $\De$ becomes larger because
of the formation of caustics which allow same-sized particles to collide with
each other with arbitrarily large velocities.

\begin{figure}[h]
\includegraphics[width=\columnwidth]{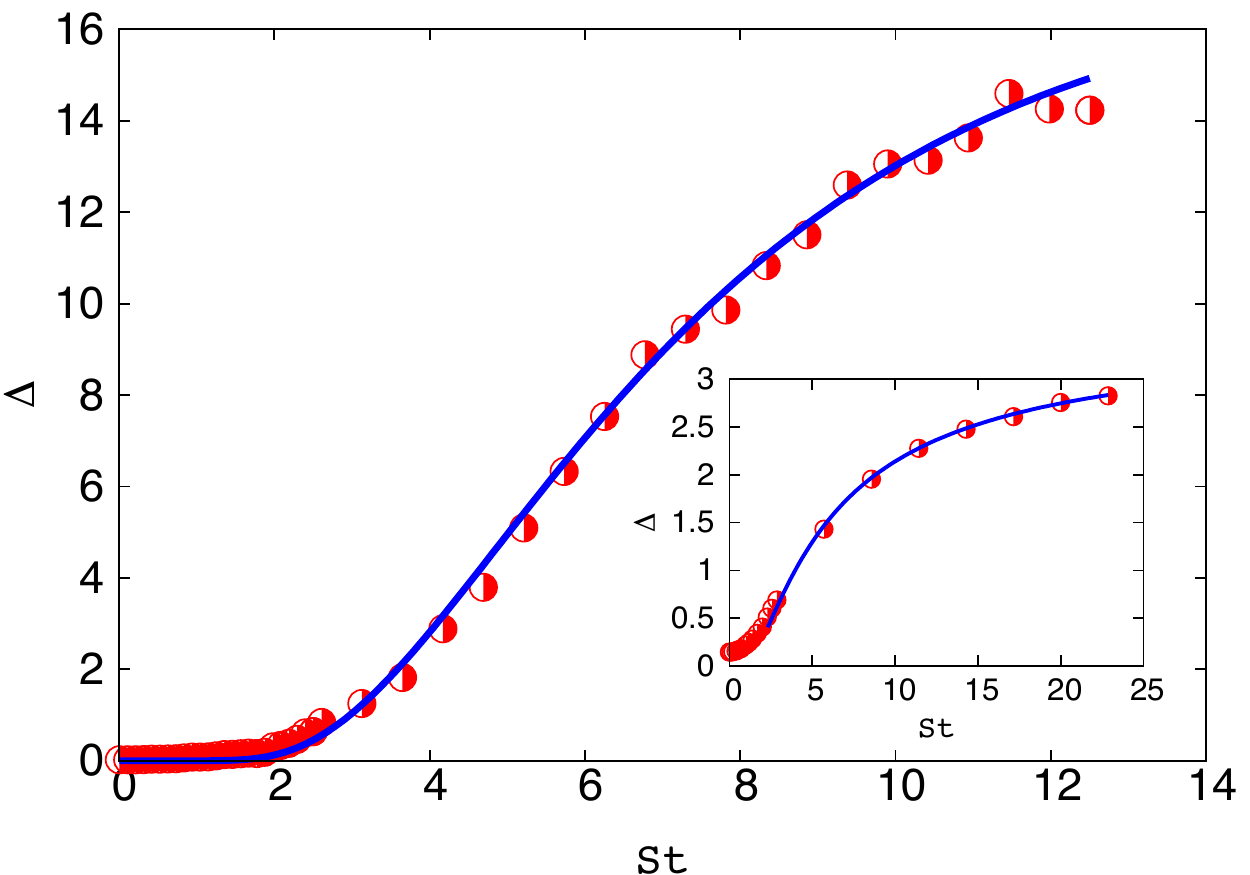}
\caption{(color online) Plot $\De$ as a function of $St$ for same-sized particles. 
The blue solid curve is 
our theoretical prediction and the symbols are data from our simulations in 2D and 3D (inset). The error bars
on our numerical data are comparable to the symbol size.}
\label{fig:same}
\end{figure}

In order to gain a complete understanding of the dependence of $\De$ on $St_1$ and $St_2$, it is useful 
to return to Eq.~\eqref{eq:particles}. Let us consider two particles 1 and 2 with Stokes times $\tau_1 = \tau$ and 
$\tau_2 = \alpha \tau_1 = \alpha \tau$. We consider the non-dimensional form of Eq.~\eqref{eq:particles} (by 
including factors of $\tau_\eta$) to obtain 
\begin{eqnarray} 
\frac{d{\bf v}_1}{d\tau} &=& -\frac{1}{St}[{\bf v}_1 - {\bf u}_1];\nonumber\\
\frac{d{\bf v}_2}{d\tau} &=& -\frac{1}{\alpha St}[{\bf v}_2 - {\bf u}_2].\nonumber
\end{eqnarray}
For brevity, we set $\bu(\xp,t) = {\bf u}_{\rm i}$. Thence we obtain 
\begin{equation}
\frac{d\De}{dt} = -\frac{1}{\alpha St}\left [\De - St(1 - \alpha)\frac{d {\bf v}_1}{dt}\cdot \hat{\bf r}\right ]. 
\label{eq:main}
\end{equation}
We have assumed here that at small particle separations $|{\bf r}|$, the fluid
velocity is smooth and hence $|{\bf u}_1 - {\bf u}_2| \equiv {\bf \sigma}\cdot
{\bf r} \sim 0$, as $|{\bf r}| \to 0$ (${\bf \sigma}$ is the gradient of the
fluid velocity).

So far we have made only one defensible assumption in deriving
Eq.~\eqref{eq:main} which has to do with the smoothness of the velocity field
at small scales. To this extent Eq.~\eqref{eq:main} is exact in both 2D and 3D.
Let us now explore the various asymptotics of this equation and obtain
theoretical estimates of $\De$ for different combinations of Stokes numbers
which can then be tested against data from our DNSs in 2D and 3D.  
The following limits naturally arise in this case (see Table~\ref{cases} for a compact version 
of these limits): Case 1: $\alpha = 1$ with no assumption 
on $St$; Case 2: Particle 1 is very small 
and close to being a tracer ($St \ll 1$) and $\alpha \sim \mathcal{O}(1)$ such that $St_2 \lesssim 1$; 
Case 3: Particle 1 is still small ($St \ll 1$) but $\alpha \gg 1$ such that $St_2 \gg 1$;
and Case 4: For $St \gtrsim 1$. To obtain the limiting form of $\De$ from Eq.~\eqref{eq:main} 
in each such case, we assume that at time $t = 0$ the particles have come close to 
each other (without actually colliding) with a velocity difference $\De_0$ and then, 
over a time $\tau$, they touch.

For $\alpha = 1$ (Case 1), integrating Eq.~\eqref{eq:main}, we obtain $\De =
\De_0\exp{(-\tau/St)}$, where (and in what follows) $\De_0$ is the constant of
integration. In Fig.~\ref{fig:same} we test our 
theoretical prediction (solid line) against data from our DNSs (symbols) in both 
2D and 3D (inset) and find excellent agreement between the two.

\begin{figure}[h]
\includegraphics[width=\columnwidth]{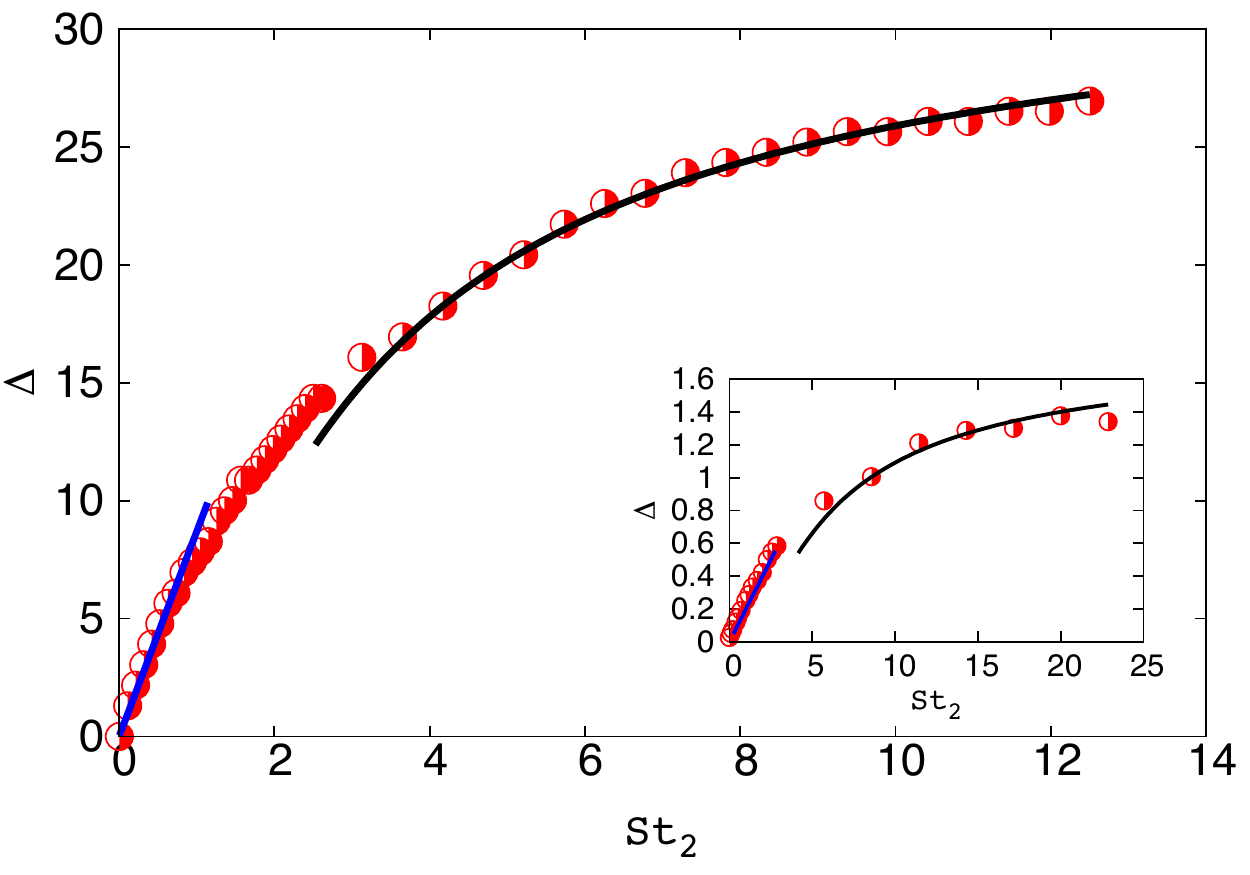}
\caption{(color online) Representative plot of $\De$, shown with symbols from our DNS data, as a function of 
$St_2$ for $St = 0.005 \ll 1$ in 2D and for tracers in 3D (inset). Our theoretical prediction for 
$St_2 < 1$ and for $St_2 \gtrsim 1$ are shown by the blue and black solid curves, respectively. 
The error bars on our numerical data are comparable to the symbol size.}
\label{fig:s_l}
\end{figure}

We now address the important question of what happens when the two colliding particles have 
different Stokes numbers. Let us begin with Case 2 where $St \ll 1$ ($v_1 \approx u_1$) 
and $\alpha \sim \mathcal{O}(1)$. In this limit, we can rewrite Eq.~\eqref{eq:main} as 
\begin{equation}
\alpha St \frac{d\De}{dt} = -\left [\De - St(1 - \alpha)\frac{d {\bf v}_1}{dt}\cdot \hat{\bf r}\right ]. 
\label{eq:main1}
\end{equation}
Since $\alpha St < 1$, we can set $\alpha St \frac{d\De}{dt} = 0$ and obtain $\De \sim St_2$.
In the other limit, Case 3, where $St \ll 1$ and $\alpha \gg 1$ 
such that  $\alpha St > 1$, we notice that 
$\frac{St(1 - \alpha)}{\alpha St}\frac{d {\bf v}_1}{dt}\cdot \hat{\bf r} \sim \frac{d {\bf v}_1}{dt}\cdot \hat{\bf r} 
\sim 0$ to leading order since $St \ll 1$. Thence we obtain $\De = \De_0\exp{(-\tau/St_2)}$.

Given the strong assumptions made in arriving at the two limits above, it is important to check our prediction 
against data from our DNSs. In Fig.~\ref{fig:s_l} we show a representative plot of the impact velocity $\De$ between 
particles of Stokes number $St = 0.005$ (2D) and tracers (3D, inset)~\cite{foot2} with particles of 
different Stokes numbers. We immediately notice that when $St_2 \lesssim 1$, $\De$ is indeed linear with 
$St_2$ and the data (symbols) consistent with our theoretical prediction shown as a blue curve. In the other limit 
when $St_2 \gg 1$, our data from numerical simulations is in excellent agreement with the theoretical 
prediction $\De \sim \exp{(-1/St_2)}$. 

At this stage it is important to remark about the constant of integration. $\De_0$ is the typically 
velocity difference with which two droplets come near each other. From very general 
conditions, it is likely that $\De_0$ should depend on turbulent intensity, the spatial dimension, as well 
as the relative Stokes numbers of the fluid (when $St_1, St_2 \sim \mathcal O(1)$. 
However we do not have an analytical expression 
for $\De_0$ and as our theoretical and numerical results suggest, $\De_0$ is likely to be a 
constant or an algebraic, sub-dominant prefactor  atleast in the range of Stokes numbers studied here~\cite{foot3}.

\begin{figure}[h]
\includegraphics[width=\columnwidth]{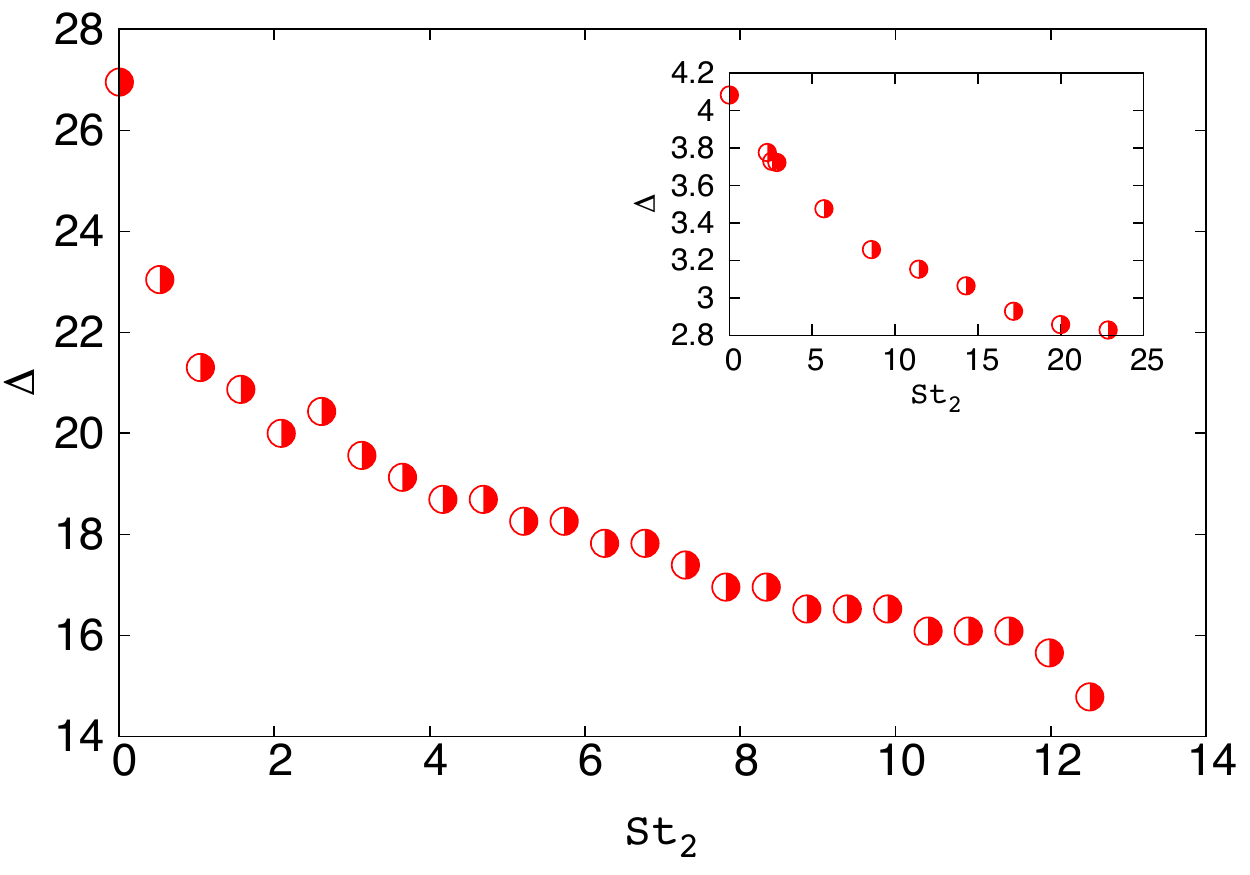}
\caption{(color online) Representative plot of $\De$ for $St = 12.5 \gg 1$ in 2D and for $St = 20 \gg 1$ 
in 3D (inset) from our numerical simulations (symbols) as a function of $St_\mathrm{2}$.
The error bars on our numerical data are comparable to the symbol size.}
\label{fig:l_s}
\end{figure}

Let us finally turn to the situation when $St \gtrsim 1$ (Case 4).  In
Fig.~\ref{fig:l_s} we show a representative plot of the impact velocity between
a particle of Stokes number $St = 12.5$ (2D) and $St = 20$ (3D, inset) with all
other particles. From our numerical data we see that that $\De$ shows rapid
variation when $St_2 \lesssim St_1$. However we do not have a self-consistent 
understanding of the functional form of the decrease in $\De$ from Eq.~\eqref{eq:main}. 

Before concluding, we return to theoretical considerations which are central to
this study. In recent years much work has gone into understanding how large
particle aggregates are formed -- mainly through coalescence -- from small
nuclei particles carried in a turbulent flow. This question is of fundamental
importance when tackling issues as diverse as formation of large rain droplets
in warm clouds, pollutant dispersion and concentration, and the growth of
planetesimals in astrophysics. A stumbling block in a fully self-consistent
theory to explain such accelerated growths is the lack of realistic
collision-coalescence kernels for different-sized particles which incorporate
both fragmentation and coalescence. A first step in this direction is, of
course, determining the dependence of impact velocities of colliding particles
on their sizes. In this paper we show, through theory and simulations, what
this dependence is.

Our results, summarised in Table~\ref{cases}, are remarkable in their implication. We
show that the larger particles (large Stokes numbers) do not collide with
arbitrarily large velocities with the smaller, tracer-like particles but
actually saturate ($\sim \exp{(-1/St)}$). This suggests that in inhomogeneous
suspensions, such as the polydisperse droplet distribution in warm clouds, a
run-away growth for large droplets through coalescence (and not fragmenting
because of large velocity differences) is likely to be the dominant mechanism
triggering rain~\cite{coal,wilkinson}.  

In conclusion, we have developed a systematic theory, validated through
detailed numerical simulations, for the impact velocity of colliding droplets
of different sizes. Remarkably, our results seem to be 
independent of dimension. In particular we have shown that there is a limiting form
for the impact velocity and hence in natural settings coalescence -- and not
fragmentation due to large $\De$ -- should be the dominant mechanism. Therefore
this work is a significant step in developing models for coalescing droplets.
Important questions related to Reynolds and Froude number effects is beyond the scope 
of the present work and the issue of collision frequencies in polydisperse suspensions is  
addressed elsewhere~\cite{martin}.  

\begin{acknowledgements}
We are grateful to R. Pandit and A. Sivakumar 
for many useful discussions and encouragement.  MJ thanks DST (India)
for support.  SSR acknowledges the
support of the DAE, Indo--French Center for Applied Mathematics (IFCAM) and the 
Airbus Group Corporate Foundation Chair in Mathematics of Complex Systems established in ICTS.
\end{acknowledgements}


\begin{thebibliography}{10}
\bibitem{foot1}  We note that the same definition can be arrived at by 
considering the velocity longitudinal structure function calculated in the limit 
where the pair separation goes to 0.
\bibitem{foot2} For the 3D simulations, we choose tracers for the reference particle; our theoretical 
predictions can be shown to hold in the case when $St_1 = 0$ as well.
\end{thebibliography}
\end{document}